\documentclass[%
 reprint,
superscriptaddress,
 amsmath,amssymb,
 aps,
 prl,
]{revtex4-2}

\usepackage{graphicx}
\usepackage{dcolumn}
\usepackage{bm}
\usepackage{xcolor}
\usepackage[hidelinks]{hyperref}

\begin{document}

\preprint{APS/123-QED}

\title{Growth and remodeling control shape memory in morphogenetic rods}

\author{Nicolas Romeo}
 \email{nromeo@uchicago.edu}
\affiliation{%
 Center for Living Systems, 
  University of Chicago, IL, USA
}%
\affiliation{%
 Department of Physics, University of Chicago, IL, USA
}%
\affiliation{%
 Department of Molecular Genetics and Cell Biology, University of Chicago, Chicago, IL, USA
 }%
\affiliation{NSF-Simons National Institute for Theory and Mathematics in Biology, Chicago, IL, USA}

\author{David B. Br\"uckner}
\affiliation{Biozentrum, University of Basel, Switzerland}
\affiliation{Department of Physics, University of Basel, Switzerland}

\author{Noah P. Mitchell}%
 \email{npmitchell@uchicago.edu}
\affiliation{%
 Department of Molecular Genetics and Cell Biology, University of Chicago, Chicago, IL, USA
 }%
\affiliation{%
 Institute for Biophysical Dynamics, University of Chicago, IL, USA
}%
\affiliation{%
 Center for Living Systems, 
  University of Chicago, IL, USA
}%
\affiliation{NSF-Simons National Institute for Theory and Mathematics in Biology, Chicago, IL, USA}

\date{\today}

\begin{abstract}

Mechanical instabilities provide a general design principle for shaping developing organs and engineering soft materials. 
However, in slender structures, simple elastic buckling tends to erase rather than preserve shape complexity: structures relax to the simplest possible shape, erasing finer detail.
Living systems nonetheless build complex, reproducible morphologies from continually remodeling material, while remaining robust to noise arising across scales.
Using analytical theory and numerical simulations of a minimal model of growing visco-elasto-plastic rods, we show that remodeling plays two opposing roles: 
At low plasticity, patterns coarsen through elastic relaxation, while high plasticity converts fluctuations into geometric disorder. This sets a trade-off between shape complexity and reproducibility with an optimal intermediate plasticity, which protects initial patterns.
Growth breaks this trade-off by suppressing both failure modes, enabling complex shapes to be reproducibly generated. Our results identify remodeling and growth rates as two knobs governing whether an encoded pattern is remembered, degraded, or transformed.
\end{abstract}

\maketitle

Living and non-living materials grow, bend, and remodel their structure to establish their shape, as seen in organs such as the gut~\cite{savin_growth_2011, shyer_villification_2013, capolupo_tissue_2026}, plants~\cite{gerbode_how_2012,moulton_multiscale_2020, armon_geometry_2011}, inflatable or swelling structures~\cite{siefert_programming_2019, pezzulla_morphing_2015}, and flexible electronics~\cite{sun_controlled_2006} alike. 
Both natural and engineered systems exploit the nonlinear dynamics induced by their slender geometry~\cite{audoly_elasticity_2018, antman_nonlinear_2005,sydney_gladman_biomimetic_2016, van_rees_growth_2017, boley_shape-shifting_2019,klein_shaping_2007, armon_geometry_2011, zhang_isometric_2026} to obtain complex morphological outcomes, often operating in the presence of fluctuations~\cite{gov_cytoskeleton_2003, kosmrlj_statistical_2017,radja_pollen_2019, jackson_scaling_2023}.
When growth is slow, morphogenesis proceeds quasi-statically and geometric effects dominate~\cite{ambrosi_growth_2019, goriely_mathematics_2017,moulton_morphoelastic_2013, alber_model_2025}. However, soft and living systems can remodel (or flow) on the same timescale as they grow, inducing dynamical effects that can substantially alter morphologies~\cite{chopin_dynamic_2017, kodio_dynamic_2020, shivers_renormalized_2026, matoz-fernandez_wrinkle_2020,al-izzi_hydro-osmotic_2018,liu_morphological_2018}. 

Morphological dynamics are especially important in biological development, which needs to navigate a fundamental tension between growth and noise. 
Morphogenesis emerges from the self-organization of cells working off patterned instructions~\cite{singh_zebrafish_2015, corson_self-organized_2017,slavkov_morphogenesis_2018}, but is executed under intrinsically stochastic conditions~\cite{tsimring_noise_2014, romeo_information_2026}. 
In particular, freed from detailed balance, biological systems can tune their rheology and growth patterns to enable robust morphogenesis~\cite{mongera_fluid--solid_2018, petridou_rigidity_2021,raspopovic_digit_2014}.
This raises a central question: How do geometric nonlinearities, rheology, and growth contribute to enable reproducible morphology starting from an initial patterned shape?

\begin{figure}[t]
    \centering
    \includegraphics[scale=0.95]{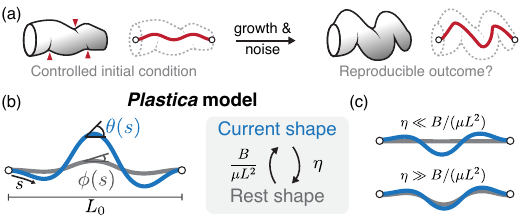}
    \caption{Does complex rheology stabilize noisy morphogenesis? (a) Consider an initially patterned growing 1D system. What conditions lead to reproducibly patterned outcomes? (b) We consider a minimal model of a growing elastica with internal remodeling rate $\eta$. The characteristic elastic rate is $B/(\mu L^2)$. (c) At low plasticity $\eta \ll  B/(\mu L^2)$, the system behaves elastically, while at large $\eta$ the system is fluid-like.}
    \label{fig:elastica1}
\end{figure}

Concrete examples of ``morphodynamic memory"
arise across \emph{Drosophila} morphogenesis. 
During germ band extension, friction between embryo and eggshell counteracts an intrinsic chiral bias~\cite{serafini_embryo-eggshell_2026},
while tension buffering suppresses fluctuation-induced buckling~\cite{smits_maintaining_2023}.
In the embryonic midgut~\cite{mitchell_visceral_2022}, early constrictions and left--right asymmetries provide spatially patterned mechanical inputs whose transformation into stereotyped loops depends on subsequent growth, confinement, and tissue remodeling~\cite{placeholder}. Understanding the (de)stabilizing effects of properties such as growth and remodeling could help control morphogenesis of living and non-living systems alike~(Fig.~\ref{fig:elastica1}a).

To make progress, we study a minimal \emph{plastica} model of visco-plastic or `morphodynamic' rods~\cite{goldstein_dynamic_2006}, which couples a one dimensional (1D) elastica to remodeling dynamics. 
Internal viscosity renders the dynamics overdamped, while plastic remodeling --- called visco-elasticity in mechanics~\cite{howell_applied_2001} --- lets the `rest' stress-free shape adapt to the current configuration at a rate $\eta$, slowly forgetting its original form.
We use this model as an analytically tractable testbed to study geometric information transmission: we prescribe an initial shape and ask how faithfully it is remembered during simulated development.

Our main finding is that remodeling is double-edged.
In the absence of plasticity or growth, complex shapes relax to the fundamental buckling mode, a process that we refer to as \textit{elastic coarsening}. Modest plasticity suppresses this effect and thereby protects complex shapes; but past a second threshold, plasticity degrades shape memory by converting noise into geometric disorder. Growth suppresses both coarsening and geometric disorder, and in doing so allows complexity and reproducibility to coexist.

\paragraph*{Plastica.} Inspired by computational models of viscous thread mechanics~\cite{bergou_discrete_2010,audoly_discrete_2013}, we consider a viscoelastic variant of the elastica: the tangent angle $\theta(s,t)$ is subject to overdamped internal viscosity $\mu$, and the bending moment \smash{$M=B(\theta'-\phi')$} is proportional to the deviation from the rest curvature $\phi' \equiv \partial_s \phi$, where $s$ is the arclength parameter (Fig.~\ref{fig:elastica1}b). 
Under an imposed deformation, the rest curvature relaxes linearly to the current configuration at a rate $\eta$, causing the bending moment to vanish at long times. Varying $\eta$, the \emph{plastica} interpolates between fully elastic (\smash{$\eta=0$}) and viscous-thread (\smash{$\eta=\infty$}) behavior~\cite{brun_liquid_2015, audoly_discrete_2013, kamrin_soft_2012}.
We consider a rod of length $L$ with pinned (simply-supported) ends spaced apart by \smash{$L_0 < L$}. Under a small-angle approximation where \smash{$\Delta \equiv L-L_0 \ll L_0$}, the plastica obeys
\begin{subequations}
    \begin{align}
        \mu \dot{\theta} & = B(\theta'' - \phi'') + F\theta + \sqrt{2\sigma}\zeta(s,t) \\
         \dot{\phi}' & = \eta (\theta' - \phi')
    \end{align} \label{eq:plastica}%
\end{subequations}%
subject to the reduced end-shortening constraint \smash{$(1/2)\int_0^{L}\mathrm{d}s\, \theta^2 = L-L_0 \equiv \Delta$},
with the shorthand \smash{$\dot{\theta} \equiv \partial_t \theta$} and standard Brownian noise \smash{$\langle \zeta(s,t)\rangle = 0$}, \smash{$\langle \zeta(s_1,t_1) \zeta(s_2,t_2)\rangle = \delta(s_1-s_2)\delta(t_1 -t_2)$}.
While Eq.~\eqref{eq:plastica} appears linear, the tension $F$ is a Lagrange multiplier imposing the quadratic constraint, making the problem nonlinear. 
We derive Eq.~\eqref{eq:plastica} from the fully nonlinear formulation valid for all $\Delta/L_0$ in~\cite{supp}; these agree in the $\Delta < 0.1L_0$ regime that follows.

The Gaussian noise here is a minimal stand-in for dynamical perturbations of the surrounding of the rod or its internal mechanics, which could be due to thermal noise or, for living systems, stochastic contractility or gene expression of proteins that set the mechanics of the system, such as filaments or motors at moderate copy counts~\cite{martin_pulsed_2009,mitchell_visceral_2022}.

To obtain a compact description of the dynamics, we expand the tangent profile in Fourier modes \smash[t]{$\theta(s,t) = \sum_{n=1}^d \theta_n(t)\cos(q_n s)$} with \smash{$q_n = \pi n/L$} and $d$ the total number of modes. For 3D materials, $d$ is set by the aspect ratio of the rod, \smash{$d \sim L_0 /h$}, with the precise form depending on the microscopic structure of the rod: thinner rods can support more modes.
In mode space the dynamics become~\cite{supp}
\begin{subequations}
    \begin{align}
        \mu \dot{\theta}_n & = (-Bq_n^2 +F)\theta_n + Bq_n^2 \phi_n+ \sqrt{2\sigma/L}\zeta_n(t) \\
         \dot{\phi}_n & = \eta (\theta_n - \phi_n)
    \end{align}\label{eq:mode_dyn}%
\end{subequations}
with the end-extension constraint now reading $\sum_{n=1}^d \theta_n^2 = 4\Delta/L \equiv C$ and $\langle \zeta_n(t) \zeta_m(t')\rangle = \delta_{nm} \delta(t-t')$. The noise amplitude is length dependent in mode space: modes of a longer rod are less affected by noise as their dynamics are averaged over more microscopic kicks.

The behavior of solutions of Eq.~\eqref{eq:mode_dyn} depends on a handful of dimensionless numbers, but two matter most in what follows. Defining the elastic timescale \smash{$\tau_E = \mu L_0^2/B$}, the plasticity number \smash{$\mathrm{Pl} = \eta \tau_E$} sets how fast the rod remodels relative to the speed of elastic relaxation (Fig.~\ref{fig:elastica1}c), while a normalized growth rate $g\tau_E$ (introduced below) sets the speed of elongation relative to elasticity. 
The remaining groups -- including the confinement ratio $\Delta/L_0$, the number of modes $d$, and a noise scale  $\bar{\sigma}= \sigma/(L_0\tau_E)$ -- are held fixed unless noted. Our goal is to understand how plasticity and growth govern the fate of an initial pattern in the presence of noise.

\begin{figure*}
    \centering
    \includegraphics[scale=0.99]{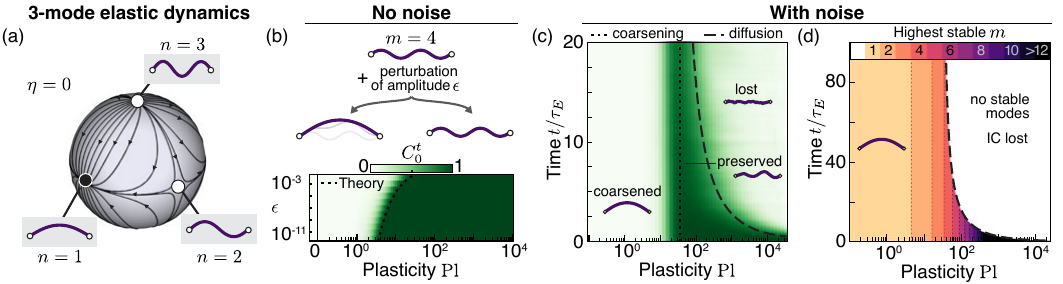}
    \caption{Plasticity suppresses deterministic coarsening but induces noise-driven state diffusion. (a) In the absence of growth, weakly-nonlinear dynamics of a fully-elastic system are constrained to a hypersphere, with one stable mode (black circle). Here, $d=3$. (b) In the absence of noise, a pure mode \smash{$m>1$} is robust to perturbations of amplitude $\epsilon$ only above a plasticity threshold $\mathrm{Pl}(\epsilon)$. Here \smash{$m=4$}. (c) High plasticity systems are vulnerable to stochastic diffusion, leading to a plasticity dependent memory time span. Boundaries derived from theory. (d) Our theory predicts that retention of the initial condition depends on its complexity, as measured by the order $m$ of its highest mode. For readability we crop the colorbar to \smash{$m=12$} below its maximal value $d$. For (b-d), simulations averaged over $96$ replicates, \smash{$d=64$}, \smash{$L/L_0=1.1$}. For (c-d), \smash{$\bar{\sigma} = 0.005$}.}
    \label{fig:nogrowth}
\end{figure*}

\paragraph{Elastic coarsening.} We first investigate shape dynamics in the purely elastic limit \smash{$\eta=0$} with a straight rest configuration \smash{$\phi=0$}. The constraint $\sum_{n=1}^d \theta_n^2 = C$ confines the mode dynamics to the sphere $\mathcal{S}^{d-1}$.  Recast in terms of the mode fractions $r_n = \theta_n^2/C$, the deterministic dynamics are replicator-like $\mu\dot{r}_n = 2B(\langle{q^2}\rangle_r - q_n ^2) r_n$ with $\langle q^2 \rangle_r = \sum_n q_n^2 r_n$ the mode occupancy-averaged wavenumber~\cite{supp}. Equation~\eqref{eq:mode_dyn} has two stable fixed points, $\theta_1 = \pm \sqrt{C}$: the fundamental mode. Every higher pure mode $\theta_n = \pm \sqrt{C}$ is a saddle with \smash{$n-1$} unstable directions flowing towards the lower modes (Fig.~\ref{fig:nogrowth}a). In analogy with population genetics, each mode has a ``fitness'' $f_n=-q_n^2$, and the smoothest, fittest mode competitively excludes the rest~\cite{nowak_evolutionary_2006}. Any perturbation from a fixed point will therefore flow deterministically to the fundamental mode, causing elastic coarsening. Thus, complex shapes cannot be retained in the absence of plasticity.

\paragraph*{Plasticity protects initial patterns.}

Turning on plasticity \smash{$\eta>0$} reshapes this landscape. Every configuration with \smash{$\phi_n=\theta_n$} is neutrally stable, so the entire constraint sphere \smash{$\sum_n \theta_n^2 =C$} becomes marginally stable: 
once the rest shape catches up to the current configuration, the system is frozen in the absence of drive or noise, as every configuration is equivalent. 
The consequences are intuitive: if plasticity is fast enough so that the system reaches \smash{$\phi_n=\theta_n$} before elastic coarsening completes, then the initial condition is protected. 

 We make this quantitative for a rod initially in pure mode $m$ (\smash{$\theta_m^2 = C$}, \smash{$\theta_{n\neq m} =0$}, and  \smash{$\phi_n = 0$} for all $n$), subjected to an instantaneous random perturbation of typical amplitude \smash{$\epsilon \ll 1$} in all modes.
Plastic adaptation gives \smash{$ \phi_n \approx (1- e^{-\eta t})\theta_n$} (assuming $\theta_n$ changes little over this time) 
so the tension is \smash{$F\approx Bq_m^2 (1- \phi_m/\theta_m)$} 
and the mode-fraction dynamics become~\cite{supp}
\begin{align}
    \mu \dot{r}_n = 2B\left( q_m^2 - q_n^2  \right)e^{-\eta t} r_n.
\end{align}
Mode $n$ stops growing after a few $1/\eta$ as \smash{$e^{-\eta t}\to 0$}. 
For coarsening to occur before plasticity freezes the dynamics, the mode fraction $r_1$ must grow from its initial value \smash{$\epsilon^2$} to \smash{$\sim 1$} over this time, which translates into the condition 
\begin{align}
    \ln(r_1/\epsilon^2) = \int_0^{\infty}\mathrm{d}t \, \frac{2B}{\mu}e^{-\eta t}\left(q_m^2-q_1^2\right) \leq \ln(1/\epsilon^2) \label{eq:amplification_integral}
\end{align}
to prevent coarsening. Eq.~\eqref{eq:amplification_integral} thus yields an $m$-dependent shape preservation criterion
\begin{align}
     \frac{\eta \mu L_0^2}{B} = \mathrm{Pl} \geq \pi^2 \left(L_0/L\right)^2(m^2 - 1) / \ln (1/\epsilon),\label{eq:pl_crit}
\end{align}%
implying that more complex shapes, composed of modes with higher $m$, require higher plasticity to be remembered. Simulations confirm this threshold, using the Pearson correlation $C_0^t$ between initial and final modes \smash{$C_0^t = \langle \rho_{r_n(t), r_n(0)} \rangle$} averaged over replicates~\footnote{The correlation $\rho$ is taken using the covariance over all modes} to assess shape preservation (Fig.~\ref{fig:nogrowth}b).
With noise \smash{$\sigma > 0$}, the same argument holds with $\epsilon^2$ now set by the amplitude of fluctuations $\epsilon_\sigma^2 = \sigma / (\mu B q_1^2 C)$. This again matches simulations for which the threshold plasticity \smash{$\mathrm{Pl} \approx 30$} at \smash{$\bar{\sigma} = 0.005$} (Fig.~\ref{fig:nogrowth}c). Together, this shows that initial shapes are preserved beyond a minimum plasticity.

\paragraph*{Coarsening vs noise-sensitivity trade-off determines optimal plasticity.} While a minimum plasticity is required to avoid elastic coarsening, simulations 
reveal that at large plasticity the rod again loses its initial pattern due to increasing sensitivity to noise, suggesting an optimal intermediate plasticity for memory (Fig.~\ref{fig:nogrowth}c).
Physically, a very plastic rod is nearly fluid and stochastic perturbations become rapidly baked into the rest configuration; mathematically, the system now diffuses freely on the constraint sphere.

To quantify this, we work in the large $\eta$ limit, justified since these stochastic effects are only relevant when \smash{$\mathrm{Pl} \gtrsim 30$}, large enough to stabilize deterministic coarsening.
In this regime, the rest state approximately tracks the current state  
as \smash[t]{$\phi_n(t) \approx \theta_n(t) - \eta^{-1}\dot{\theta}_n(t)$}, giving effective dynamics \smash[b]{$\gamma_n \dot{\theta}_n = F \theta_n + \sqrt{2\sigma/L}\,\zeta_n(t)$}~\cite{supp};
with a mode-dependent `viscosity' \smash{$\gamma_n = \mu + B q_n^2/\eta$}. Without growth, the spherical constraint forces \smash{$F=0$}, and the system performs an anisotropic Brownian motion on the sphere, diffusing away from the initial configuration with diffusion constant $D_d$.

\begin{figure*}
    \centering
    \includegraphics[scale=0.95]{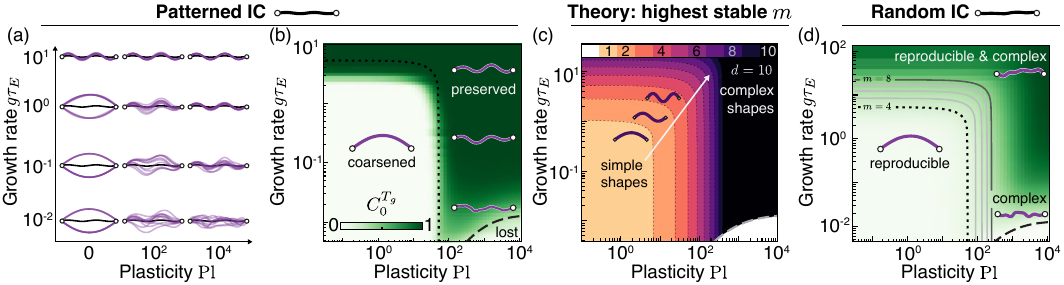}
    \caption{Growth stabilizes patterns. With an initial condition $m=4$, we find that growth and plasticity cooperate to stabilize the initial pattern. (a) Representative centerlines for 10 replicates. (b) Dotted lines indicate coarsening threshold, dashed diffusion. (c) Stabilizing increasingly complex (higher $m$) initial conditions  (dotted and gray lines) is possible with fast growth and high plasticity. (d) Theory is consistent with numerics starting from a random initial condition, highlighting conditions for reproducible growth of complex shapes. For (a-d), $L_1/L_0=1.005$, \smash{$L_f/L_1 = 1.1$}, \smash{$\bar{\sigma} = 0.001$} and \smash{$d=10$}. Results averaged over 96 replicates. See~\cite{supp} for additional simulations with varying $d$ and fully-nonlinear dynamics.}
    \label{fig:growth}
\end{figure*}

How long does it take for the system to diffuse in mode space? 
Correlations will be lost after a time \smash{$t_D\sim D_d^{-1}$}. Crucially, not all modes diffuse: high-frequency wrinkles are too stiff to be stirred by noise, and the shape diffuses in a space of lower dimension than its full mode count. Weighting each direction by its contributions to the anisotropic diffusion tensor gives $D_d = (d_\text{eff}-1) D_0$ with \smash{$d_\text{eff}(\eta) = \sum_{n=1}^d \left(1+ Bq_n^2/\mu \eta \right)^{-2}$} and $D_0 = \sigma / (\mu^2\Delta)$. As \smash{$\mathrm{Pl}\to\infty$}, \smash{$d_\mathrm{eff} \to d$} the total number of modes in the spectral representation, and diffusion becomes isotropic on the sphere; at small $\mathrm{Pl}$ only the fastest modes participate and \smash{$d_\text{eff}\to 0$}. The resulting memory time  $t_D(\mathrm{Pl})/\tau_E = (\mu/\tau_E)^2 (\Delta/L_0)/[(d_\text{eff}-1) \bar{\sigma}]$ agrees with simulations (Fig.~\ref{fig:nogrowth}c; see \cite{supp} for varying $d$ and noise amplitude $\sigma$ and steady-state distribution).

This memory time is independent of the initial pattern, while the threshold for protection from coarsening is $m$-dependent (Eq.~\eqref{eq:pl_crit}). Between the two, there is a window in which patterns are remembered, narrowing for more complex shapes with higher \smash{$m\leq d$} (Fig.~\ref{fig:nogrowth}d). Plasticity can thus protect or corrupt the memory of the initial shape, causing a trade-off between sensitivity to noise and robustness against coarsening, which depends on the complexity of the shape to be remembered.

\paragraph*{Growth breaks the trade-off.} 
A central feature of morphogenetic systems is their intrinsic growth, modeled here as uniform elongation of the rod. As the rod elongates, both the elastic coarsening rate \smash{$B/(\mu L(t)^2)$} and the spectral noise amplitude \smash{$2\sigma/L(t)$} decay. The former stems from the fact that a longer rod at a fixed mode amplitude is less curved, while the latter reflects that noise is averaged over more material. Thus, growth affects both failure modes of plasticity-based shape memory at once.

First, we consider the effect on elastic coarsening. For concreteness, we consider exponential growth \smash{$L(t) = L_1 e^{gt}$} at a rate $g$ from \smash{$L_1 > L_0$} to a final time \smash{$T_g = \ln (L_f/L_1) /g$};  other growth laws follow similarly. Evaluating the amplification integral Eq.~\eqref{eq:amplification_integral} over the time interval $[0, T_g]$
gives a growth-dependent coarsening threshold: strikingly, even when \smash{$\mathrm{Pl}\to 0$}, there exists a growth rate $g_0$ that suppresses elastic coarsening given by \smash{$g_0 \tau_E = \pi^2 (L_0/L_1)^2(m^2 - 1) (1 -  L_1^2/L_f^2)/ \ln(1/\epsilon_\sigma)$}~\cite{supp}. This is consistent with the known fact that rapid compression causes higher buckling modes to be observable~\cite{gladden_dynamic_2005}. At large plasticity, growth-induced tension opens an alternative coarsening mechanism, but it is negligible at large $\mathrm{Pl}$ and mostly shifts the deterministic coarsening boundary~\cite{supp}. Note here that $B$ stays constant in time: other constitutive laws in which the rod thins as it elongates would suppress coarsening even faster~\cite{goriely_mathematics_2017}.

Second, we consider the effect on noise sensitivity. At large plasticity, noise-driven mode mixing remains dominant, and the total drift in mode space is now given by $\Phi(T_g) = \int_0^{T_g} \mathrm{d}t\,(d_\text{eff}(t)-1) D_0(t)$, which can be explicitly evaluated and must remain \smash{$\leq 1$} for proper shape memory.
In the large $\eta$ limit, \smash{$d_\text{eff} \to d$} from below and the integral converges, giving a critical growth rate \smash{$g_\infty \propto (d-1)\bar{\sigma}/\tau_E$} above which noise can no longer erase the pattern. Explicit expressions for $g_\infty$ and $\Phi(T_g)$ are provided in the supplement~\cite{supp}. 
Both criteria are consistent with simulations starting from a single mode $m=4$ (Fig.~\ref{fig:growth}a-b).

This analysis yields a simple design rule: to stabilize arbitrary initial conditions, there is an optimal quadrant of high growth and high plasticity. At slow growth, the plasticity trade-off is relevant: reproducible but simple shapes at low $\mathrm{Pl}$, or complex but variable shapes at high $\mathrm{Pl}$. Fast growth, however, enables reproducibility and complexity to go hand in hand (Fig.~\ref{fig:growth}c-d).

\paragraph*{Discussion.}

Here, we studied the influence of growth and plastic adaptation on the ability of a slender structure to preserve or transform an initially encoded pattern during development.
In this formulation, the relevant question is not simply whether a tissue reaches a final shape, but whether the system retains the intended geometric program despite noise. Our minimal model gives a clear answer: remodeling protects patterns at modest rates but corrupts them above a higher threshold. Growth nullifies this trade-off between complexity and reproducibility.

This framework is relevant to systems where rheology tunes instabilities to set shape~\cite{nelson_buckling_2016, ingber_mechanical_2006,petridou_tissue_2019}, and to the design of fluctuation-tolerant compliant materials.
For example, in the \emph{Drosophila} midgut, the results give a concrete prediction: an early constriction is amplified into a stereotyped loop only within a window of remodeling rates -- fast enough that elastic relaxation does not take place, slow enough that noise is not integrated. Rapid growth should widen that window, making increasingly intricate patterns reproducible. Similar  processes could be relevant for folded surfaces such as the brain's~\cite{tallinen_growth_2016,choi_biophysical_2025}.

Our analysis here is restricted to a single rod with constant material properties; other dynamics can lead to more complex behavior, with external viscosity (rather than internal) already leading to intermediate coarsening~\cite{supp}. Further work could explore similar effects in surfaces~\cite{shivers_renormalized_2026},  optimal control and patterning of spatial properties~\cite{mongera_fluid--solid_2018,hoffmann_grow_2026}, or bundles of growing and remodeling rods~\cite{imran_alsous_physical_2026, tatulea-codrean_elastohydrodynamic_2022}.
Finally, the mapping between buckling mode-competition and replicator dynamics~\cite{schuster_replicator_1983, cressman_replicator_2014} suggests that the outcome of the interplay of elasticity, remodeling, growth, and noise found here may be relevant wherever competing components are regulated by adaptive feedback~\cite{stern_physical_2025}.

\begin{acknowledgments}

\paragraph*{Acknowledgments.} We thank Brato Chakrabarti and Ousmane Kodio for insightful discussions and Cal Floyd for comments.  
This research was supported by NICHD R00HD110675, grants from the NSF (DMS-2235451) and Simons Foundation (MPS-NITMB-00005320) to the NSF-Simons National Institute for Theory and Mathematics in Biology (NITMB), and the Physics Frontier Center for Living Systems funded by the National Science Foundation (PHY-2317138). 
NR acknowledges support from the University of Chicago Center for Living Systems Fellowship, with partial support from the University of Chicago AI Initiative and The Knapp Center.
Computing resources were provided by the University of Chicago Research Computing Center. Claude (Anthropic) was used to assist with simulations and mathematical drafting. Code is available~\footnote{Code is available at \url{https://github.com/MitchellLabCode/plastica}}. 
\end{acknowledgments}

\bibliography{biblio}

\end{document}


\preprint{APS/123-QED}

\author{Nicolas Romeo}
 \email{nromeo@uchicago.edu}
\affiliation{%
 Center for Living Systems, 
  University of Chicago, IL, USA
}%
\affiliation{%
 Department of Physics, University of Chicago, IL, USA
}%
\affiliation{%
 Department of Molecular Genetics and Cell Biology, University of Chicago, Chicago, IL, USA
 }%
\affiliation{NSF-Simons National Institute for Theory and Mathematics in Biology, Chicago, IL, USA}

\author{David B. Br\"uckner}
\affiliation{Biozentrum, University of Basel, Switzerland}
\affiliation{Department of Physics, University of Basel, Switzerland}

\author{Noah P. Mitchell}%
 \email{npmitchell@uchicago.edu}
\affiliation{%
 Department of Molecular Genetics and Cell Biology, University of Chicago, Chicago, IL, USA
 }%
\affiliation{%
 Institute for Biophysical Dynamics, University of Chicago, IL, USA
}%
\affiliation{%
 Center for Living Systems, 
  University of Chicago, IL, USA
}%
\affiliation{NSF-Simons National Institute for Theory and Mathematics in Biology, Chicago, IL, USA}

\title{Supplement to: Growth and remodeling control shape memory in morphogenetic rods}

\vspace{-20pt}

\maketitle

\vspace{-30pt}
\tableofcontents

\section{Plastica}

\subsection{Fully-nonlinear formulation}

We consider the elastica system, with a constitutive equation $M(s,t) = B\left(\theta'(s,t) - \phi'(s,t)\right)$, where $\theta$ is defined by the tangent vector to the curve $\mathbf{t}(s) = (\cos\theta(s), \sin \theta(s))^T$ and $B$ is the bending modulus of the rod. Throughout, we will use dots ($\dot{\circ}$) to denote time-derivatives and primes ($\circ'$) for derivatives with respect to the curve parameter $s$.
To model internal viscosity, we consider that the material viscosity induces an orientational relaxation timescale that dominates over the linear momentum dissipation timescale set by the viscosity of an external medium, such that the internal linear and angular momentum balance reads  
\begin{subequations}
    \begin{align}
        \mathbf{F}' & = 0 \label{eq:tension_0}\\
        M' + F_x \sin\theta  - F_y \cos\theta & = \mu \frac{D\theta}{Dt}
    \end{align}
\end{subequations}
where $D/Dt$ denotes the material derivative. As a consequence of the linear momentum balance Eq.~\eqref{eq:tension_0}, the tension of the rod $\mathbf{F} = ( F_x, F_y)^\top$ is constant throughout the rod with no external viscosity.
To model plasticity of the material, we append a linear update rule for $\phi$ such that the reference curvature $\phi'$ tracks the current configuration with a timescale $\eta$, which reads
\begin{align}
    \frac{D\phi'}{Dt} = - \eta(\phi' - \theta').
\end{align}
We consider the situation where the ends are pinned and aligned, which imposes the two constraints
$\int_0^L \mathrm{d}s\, \cos\theta = L_0$ and $\int_0^L \mathrm{d}s\, \sin\theta = 0$. These constraints set the values of $F_x$ and$F_y$ which lead to the angle dynamics.
\begin{align}
    \mu  \frac{D\theta}{Dt} & = B ( \theta'' - \phi'') + F_x \sin \theta - F_y \cos \theta \\
    \frac{D\phi'}{Dt} & = \eta( \theta'- \phi')
\end{align}
with $D/Dt = \partial_t  + \dot{s}\partial_s = \partial_t  + gs\partial_s $ for homogeneous exponential growth with $g = \dot{L}/L$.
With pinned moment-free boundary conditions such that $\theta'=\phi'$ at both ends (and $\phi(0)=0$),  we can expand on cosine modes, with $q_n = \pi n/ L(t)$
\begin{align}
    \theta(s,t) = \sum_{n=1}^d \theta_n(t) \cos(q_n(t) s), \quad \phi(s,t) = \sum_{n=1}^d \phi_n(t) \cos(q_n(t) s).
\end{align}
$d$ is the total number of modes retained in the spectral truncation, set by the physical length scale setting the validity of the one-dimensional dynamics. In mode space, we then have, accounting for the convective correction $gs \partial_s \theta$ in the presence of homogeneous growth with $g = \dot{L}/L$,
\begin{align}
    \mu \dot{\theta}_n & = B q_n^2(\phi_n-\theta_n) + \mathcal{F}_n[F_x \sin\theta - F_y \cos\theta] \\
    \dot{\phi}_n & = \eta( \theta_n -\phi_n)
\end{align}
with the inverse cosine transform $\mathcal{F}_n[f] = \frac{2}{L}\int \mathrm{d}s\, f(s)\cos(q_ns)$.

\subsection{Small-angle approximation of the plastica}

Assuming $|\theta| \ll 1$, by expanding the trigonometric functions to first order we can then reduce the equations to 
\begin{align}
    \mu \dot{\theta}_n & = (-B q_n^2 + F)\theta_n + B q_n^2 \phi_n \label{eq:modedyn_1}\\
    \dot{\phi}_n & = \eta( \theta_n -\phi_n)
\end{align}
with $F_y=0$ at this order with the moment-free boundary conditions while $F = F_x$ denotes the remaining component of the tension.
In the linearized regime, the inextensibility constraint simplifies to a quadratic  $L_0  \approx L(t) - (1/2)\int_0^L \mathrm{d}s \,\theta^2$, and reads in terms of modes as 
\begin{align}
    \sum_{n=1}^d \theta_n^2 = 4 \Delta(t)/ L(t) \equiv C
\end{align}
with $\Delta(t) = L(t) - L_0 >0$ the effective end-shortening. We can then find $F$ by differentiating the constraint in time which leads to
\begin{align}
    F = \frac{\mu\dot{C}}{2C} + \frac{B}{C}\sum_{n=1}^d  q_n^2 \theta_n (\theta_n - \phi_n).\label{eq:tension}
\end{align}

\subsection{Numerical implementation}

We implement \texttt{julia} simulations to numerically solve the plastica equations, using different schemes for linear and nonlinear formulations. 

\subsubsection{Small-angle formulation}

We use a spectral semi-implicit scheme to solve the linearized scheme:  the tension is computed explicitly, but the timestepping of the ODEs is done implicitly. More precisely, at timestep $k$ we compute $T_k = \mu \dot{C}(t_k)/C(t_k) + \frac{B}{C(t_k)}\sum_n \theta_n(t_k)(\theta_n(t_k)-\phi_n(t_k))$, then solve for the angles as 
\begin{align}
    \mu(\theta_n^{k+1}-\theta_n^k) & = (- Bq_n^2 + T_k) \theta_n^{k+1} \Delta t + B q_n^2 \phi_n^{k+1} \Delta t + \zeta_n(t_k) \\
    \phi_n^{k+1}-\phi_n^k & = \eta (\theta^{k+1}_n - \phi_n^{k+1}) \Delta t
\end{align}
which can be recast as
\begin{align}
    \begin{pmatrix}
        \mu + (Bq_n^2-T_k)\Delta t &  - B q_n^2 \Delta t\\ -\eta \Delta t & 1+\eta \Delta t 
    \end{pmatrix} \begin{pmatrix}
        \theta^{k+1}_n \\\phi_n^{k+1}
    \end{pmatrix}  = \begin{pmatrix}
        \mu \theta^{k}_n + \zeta_n(t_k) \\\phi_n^{k}
    \end{pmatrix}
\end{align}
where $\zeta_n = \sqrt{2\sigma \Delta t/L}\epsilon_n$, with $\epsilon_n \sim N(0,1)$. We analytically compute the matrix inverse at each time to obtain 
\begin{align}
        \begin{pmatrix}
            \theta^{k+1}_n \\\phi_n^{k+1}
        \end{pmatrix}  = \frac{1}{D_n}\begin{pmatrix}
             1+\eta \Delta t  &   B q_n^2 \Delta t\\ \eta \Delta t & \mu + (Bq_n^2-T_k)\Delta t
        \end{pmatrix} \begin{pmatrix}
            \mu \theta^{k}_n + \zeta_n(t_k) \\\phi_n^{k}
        \end{pmatrix}\label{eq:update_linearized}
\end{align} 
with $D_n = [\mu + (\mu \eta + Bq_n^2)\Delta t] - T_k \Delta t( 1 + \eta \Delta t )$. Eq.~\eqref{eq:update_linearized} summarizes the time stepping loop used in simulations.

\subsubsection{Fully nonlinear formulation}

 The fully nonlinear dynamics is solved using a finite-difference approximation of the differential operators, with a nonlinear implicit Euler time stepper which requires iterative Newton solves.
The arclength domain $s\in[0, L(t)]$ is sampled at $N_\mathrm{pts}$ equispaced points
\begin{align}\label{eq:fd_grid}
    s_i = (i-1)\,\delta, \quad i = 1,\ldots,N_\mathrm{pts}, \qquad \delta = \frac{L(t)}{N_\mathrm{pts} - 1}.
\end{align}
The grid is regenerated every time-step as $L$ grows and $\delta$ varies in time but the number of nodes is fixed.
We use the standard second-order centered discrete Laplacian on interior points
\begin{align}\label{eq:fd_d2}
    (\partial_s^2 \theta)_i \approx \frac{\theta_{i-1} - 2\theta_i + \theta_{i+1}}{\delta^2}, \quad i = 2,\ldots,N_\mathrm{pts}-1.
\end{align}
while first derivatives at boundaries are evaluated using second-order one-sided stencils
\begin{align}\label{eq:fd_bc}
    (\partial_s\theta)_1 &\approx \frac{-3\theta_1 + 4\theta_2 - \theta_3}{2\delta}, \qquad
    (\partial_s\theta)_{N_\mathrm{pts}} \approx \frac{3\theta_{N_\mathrm{pts}} - 4\theta_{N_\mathrm{pts}-1} + \theta_{N_\mathrm{pts}-2}}{2\delta}.
\end{align}
These enforce the moment-free BCs $\partial_s\theta(0) = \partial_s\phi(0)$ and $\partial_s\theta(L) = \partial_s\phi(L)$ via the corresponding one-sided stencils applied to both $\theta$ and $\phi$. To evaluate the constraint integral constraints, we use a composite Simpson 1/3 quadrature (if $N_\text{pts}$ is odd)
\begin{align}\label{eq:fd_simpson}
    \int_0^L f(s)\,ds \approx \frac{\delta}{3}\left[f_1 + 4f_2 + 2f_3 + 4f_4 + \cdots + 4f_{N_\mathrm{pts}-1} + f_{N_\mathrm{pts}}\right] \quad (N_\mathrm{pts}\text{ odd}),
\end{align}
with the standard pattern $(1, 4, 2, 4, 2, \ldots, 4, 1)$, falling back to trapezoidal weights when $N_\mathrm{pts}$ is even. This provides $O(\delta^4)$ accuracy on the clamp constraints for smooth integrands.

At each timestep, we analytically solve for the updated $\phi$ field as
\begin{align}
    \phi^{n+1} = \alpha \theta^{n+1} + (1-\alpha) \phi^n \label{eq:theta0_implicit}
\end{align}
with $\alpha = \eta \Delta t/ (1+ \eta \Delta t)$, and then evaluate the Newton-Raphson residual for 
the state vector $\mathbf{u} = (\theta_1, \ldots, \theta_{N_\mathrm{pts}}, F_x, F_y)\in\mathbb{R}^{N_\mathrm{pts}+2}$.  Defining $\bar{B} \equiv B(1-\alpha)/\delta^2$ and using~\eqref{eq:fd_d2}, the residual is
\begin{subequations}\label{eq:fd_residual}
\begin{align}
    R_i^\mathrm{interior} &= \mu\frac{\theta_i^{n+1} - \theta_i^n}{\Delta t} - \bar{B}\left[(\theta_{i-1} - 2\theta_i + \theta_{i+1})^{n+1} - (\phi_{i-1} - 2\phi_i + \phi_{i+1})^n\right] \notag\\
    &\quad - F_x\sin\theta_i^{n+1} + F_y\cos\theta_i^{n+1}, \quad i = 2,\ldots,N_\mathrm{pts}-1,\\
    R_1^\mathrm{BC} &= \frac{-3\theta_1^{n+1} + 4\theta_2^{n+1} - \theta_3^{n+1}}{2\delta} - \frac{-3\phi_{1}^n + 4\phi_{2}^n - \phi_{3}^n}{2\delta},\\
    R_{N_\mathrm{pts}}^\mathrm{BC} &= \frac{3\theta_{N_\mathrm{pts}}^{n+1} - 4\theta_{N_\mathrm{pts}-1}^{n+1} + \theta_{N_\mathrm{pts}-2}^{n+1}}{2\delta} - \frac{3\phi_{N_\mathrm{pts}}^n - 4\phi_{N_\mathrm{pts}-1}^n + \phi_{N_\mathrm{pts}-2}^n}{2\delta},\\
    R_{F_x} &= \sum_i w_i^\mathrm{int}\cos\theta_i^{n+1} - L_0,\\
    R_{F_y} &= \sum_i w_i^\mathrm{int}\sin\theta_i^{n+1},
\end{align}
\end{subequations}
where $w_i^\mathrm{int}$ are the Simpson weights including the arclength factor $\delta/3$. Note that since the grid expands homogeneously, the material derivative is correctly accounted by the finite difference here. The Jacobian $\mathbf{J}\in\mathbb{R}^{(N_\mathrm{pts}+2)\times(N_\mathrm{pts}+2)}$ has a banded structure with off-diagonal Lagrange-multiplier coupling, which allows us to use a sparse CSC LU solver from Julia's \texttt{SparseArrays} for the increments $\mathbf{J}(\mathbf{u}^{k+1} - \mathbf{u}^k) = - R$ where $\mathbf{u}^k$ is the $k$-th iteration of the Newton step. 

\begin{figure}
    \centering
    \includegraphics[scale=1]{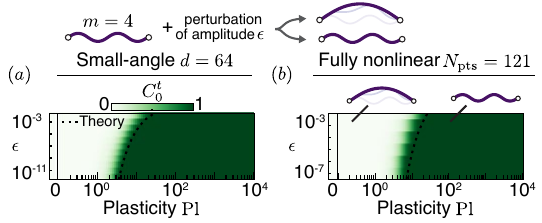}
    \caption{Small-angle and fully nonlinear models predict similar transitions for the stability of $m=4$ patterns in the absence of noise. For both (a) results are averaged over $96$ replicates, for (b) over $36$ replicates. For both, $L/L_0 = 1.1$. The range of applied perturbation amplitudes $\epsilon$ is limited in the nonlinear simulations by the accuracy of the discretization scheme.}
    \label{sfig:full_vs_linear_det}
\end{figure}

To include noise, each time step adds a Gaussian increment to $\theta$ before the Newton solve:
\begin{align}\label{eq:fd_noise_update}
    \theta_i^n \leftarrow \theta_i^n + \sigma_i\,\mathcal{N}(0,1), \qquad \sigma_i = \sqrt{\frac{2 \sigma \Delta t}{V_i}}.
\end{align} where we scale the noise amplitude to  account for the variable grid spacing $V_i$
\begin{align}\label{eq:fd_noise_volume}
    V_i = \begin{cases} \delta & i = 2,\ldots,N_\mathrm{pts}-1 \text{ (interior)} \\ \delta/2 & i = 1, N_\mathrm{pts} \text{ (boundary)} \end{cases}.
\end{align}
This scheme is consistent with the continuum variance rule $\langle \zeta(s,t)\zeta(s',t')\rangle = \delta(s-s')\delta(t-t')$.  The Newton solve then projects the noisy $\theta_i^n$ onto the constraint manifold.

In summary, each time step $n\to n+1$ includes the following steps
\begin{enumerate}
    \item Update $L^{n+1} = L_0 e^{g\,t^{n+1}}$, $\delta^{n+1} = L^{n+1}/(N_\mathrm{pts}-1)$, and the Simpson weights $w_i^\mathrm{int}$ from~\eqref{eq:fd_simpson}.
    \item Add noise $\theta_i^n \leftarrow \theta_i^n + \sigma_i\mathcal{N}(0,1)$ via~\eqref{eq:fd_noise_update}, with the new $\delta^{n+1}$ in $V_i$.
    \item Newton iterate~\eqref{eq:fd_residual} until the residual $|\mathbf{R}|_\infty < $ tolerance ($\sim 10^{-8}$);
    \item Recover $\phi^{n+1}$ from~\eqref{eq:theta0_implicit}.
\end{enumerate}
In practice, we find that this solver is consistent with the small-angle spectral solver for perturbations of amplitude $\epsilon > 10^{-6}$ in deterministic simulations (Fig.~\ref{sfig:full_vs_linear_det}). At smaller $\epsilon$, the limited precision of this solver which estimates spatial derivatives with $O(\delta^2)$ accuracy leads to divergent results with the spectrally-accurate semi-linear small-angle solver.

\subsection{Pearson correlation}

To measure preservation of the initial condition, throughout this article we use the Pearson correlation coefficient averaged over all modes. Mathematically, we define
\begin{align}
    C_0^t = \langle \rho_{r_n(t), r_n(0)} \rangle
\end{align}
with $\langle .\rangle$ the average over $N$ replicates and 
\begin{align}
    \rho_{r_n(t), r_n(0)} = \frac{1}{d}\sum_{n=1}^d \frac{z_n(t)z_n(0)}{\sigma_{z_n(t)} \sigma_{z_n(0)}}
\end{align}
the Pearson correlation between the two mode vectors, where $z_i(t)=r_n^{(i)} - d^{-1}\sum_k r_k^{(i)}$ is the mode fraction minus the mode-averaged fraction for the $i$-th replicate at time $t$ and $\sigma_{z_i(0)}$ is the per-replicate standard deviation across the mode fractions. If $r_n^{(i)}(t) = r_n^{(i)}(0)$ then $\rho_{r_n(t), r_n(0)} =1$, while if the both arguments are uncorrelated $\langle \rho_{r_n(t), r_n(0)}\rangle = 0$. Consequently, $C_0^t$ assesses the preservation of the initial condition at time $t$.

\section{Derivation of memory criteria in the absence of growth}

In this section, we derive in more detail the deterministic and stochastic criteria for shape preservation presented in the main text. We also compare the numerically-observed steady-state spectrum in the large plasticity regime to theoretical predictions.

\subsection{Derivation of replicator equations}

We first derive the dynamics of the mode fraction $r_n(t) = \theta_n^2/C$ in the absence of growth and noise. By construction, the mode fractions satisfy $\sum_{n=1}^d r_n = 1$ and their dynamics thus live on the $d$-dimensional unit simplex.
By applying the chain rule, we have
\begin{align}
    \mu\dot{r}_n = 2\frac{\theta_n \dot{\theta}_n}{C} =  2\left(F - Bq_n^2 (1-\phi_n/\theta_n)\right) r_n
\end{align}
and the tension reads, per Eq.~\eqref{eq:tension} when $\dot{C}=0$, 
\begin{align}
    \mu\dot{r}_n = 2 \left(\sum_{k=1}^d Bq_k^2r_k(1-\phi_k/\theta_k) - Bq_n^2 (1-\phi_n/\theta_n)\right) r_n.
\end{align}
Assuming that plasticity is sufficiently rapid for $\phi_n = (1-e^{-\eta t})\theta_n$ for all $n$, we then find the time-decaying replicator equation
\begin{align}
    \mu\dot{r}_n = 2 B\left(\sum_{k=1}^d q_k^2r_k - q_n^2 \right) e^{-\eta t} r_n. \label{eq:replicator_damped}
\end{align}
which reduces to standard replicator dynamics when $\eta = 0$. Here, the ``species'' are the bending modes, and $r_n=\theta_n^2/C$ is the fraction of the ``population'' in the $n$-th mode. Equation~\eqref{eq:replicator_damped} has the standard replicator form $\dot{r}_n=r_n\left(f_n-\bar{f}\right)$, with an effective fitness $f_n\propto -q_n^2$. Thus, long-wavelength, low-$n$ modes increase in relative weight at the expense of shorter-wavelength modes. This redistribution is the mechanical coarsening process. When $\eta >0$, the factor $e^{-\eta t}$ acts as a decaying selection strength, so plasticity slows and eventually arrests the redistribution.

\begin{figure}
    \centering
    \includegraphics[scale=1]{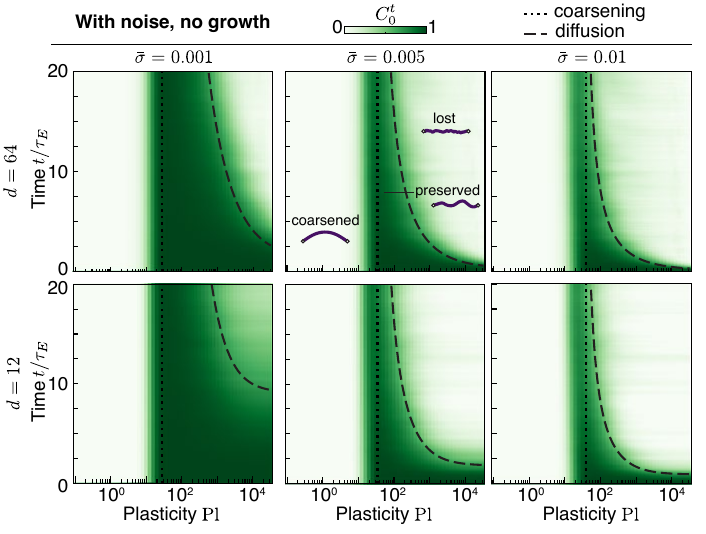}
    \caption{Theoretical regime bounds are valid across different noise amplitudes and mode numbers in the absence of growth. Results averaged over $96$ replicates; initial pattern has $m=4$, and we keep $L/L_0=1.1$. We vary $d=64$ (top row) and $d=12$ (bottom row), and noise amplitude $\bar{\sigma} \in [0.001, 0.005, 0.01]$ from left to right column. Small-angle formulation.}
    \label{sfig:varkT_d_nogrowth}
\end{figure}

\subsection{Derivation of large plasticity effective dynamics}

To study the interactions between plasticity and noise, we study the effective dynamics when plasticity is large enough to mostly wash out the memory of previous states. 
More formally, the equation $\dot{\phi}_n = \eta (\theta_n - \phi_n)$ is solved via the integral
\begin{align}
    \phi_n(t) = \phi_n(0)e^{-\eta t} + \int_0^t \frac{\mathrm{d}\tau}{\eta} e^{-\eta(t-\tau)} \theta_n(\tau).
\end{align}
When $\eta$ is large we can ignore the initial condition as $\eta t \gg 1$ and Taylor-expand using Watson's lemma
\begin{align}
    \phi_n(t) & = \int_0^t \frac{\mathrm{d}\tau}{\eta} e^{-\eta(t-\tau)} \left(\theta_n(t) + (\tau -t)\dot{\theta}_n(t) + O((\tau-t)^2)\right) \\
    & = \theta_n(t) - \frac{1}{\eta}\dot{\theta}_n(t) + O\left(\frac{1}{\eta^2}\right)
\end{align}
Inserting this relationship into Eq.~\eqref{eq:modedyn_1}, we find the effective dynamics
\begin{align}
    \gamma_n\dot{\theta}_n = F\theta_n + \sqrt{\frac{2\sigma}{L}} \zeta_n(t), \text{ with }  \gamma_n = \mu + \frac{Bq_n^2}{\eta}.
\end{align}
To find $F$, we differentiate the constraint $\sum_n \theta_n^2 = C$, which ignoring the noise, leads to
\begin{align}
    F\left(\sum_n \frac{\theta_n^2}{\mu + Bq_n^2/\eta}\right) = 0
\end{align}
which implies $F=0$ at lowest order in $1/\eta$ and noise since the sum is positive. Thus, in the absence of growth, the large-plasticity dynamics have no deterministic drift at leading order: once the rest shape tracks the current shape, every point on the constraint sphere $\sum_n \theta_n^2=C$ is marginally stable. When noise is present, the constraint projects the stochastic fluctuations onto this sphere, so the modal state undergoes the anisotropic Brownian motion described in the main text. The anisotropy is set by the mode-dependent mobility $\gamma_n^{-1}$: high-wavenumber modes have larger $\gamma_n$ and therefore diffuse more slowly, whereas $\gamma_n\to\mu$ for all $n$ as $\eta\to\infty$, yielding isotropic diffusion. This noise-driven motion away from the initial modal state is the mechanism responsible for the loss of "shape memory"  at high plasticity. Its characteristic timescale $t_D\sim D_d^{-1}$ sets the high-plasticity diffusion boundary shown in Fig. 2c of the main text and Fig.~\ref{sfig:varkT_d_nogrowth}; the dependence of $t_D$ on the total number of modes $d$ and noise amplitude $\sigma$ is consistent with numerical observations (Fig.~\ref{sfig:varkT_d_nogrowth}).

\subsection{Steady-state spectrum}

\begin{figure}
    \centering
    \includegraphics[scale=1]{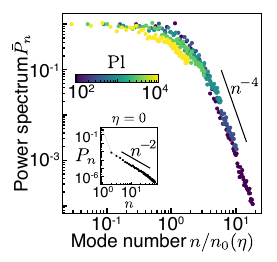}
    \caption{In the presence of noise, plastic systems converge to a late time power spectrum $\bar{P}_n = \langle \theta_n^2\rangle / \langle \theta_1^2\rangle$, with higher plasticity leading to uniform spectral distributions. Inset shows the late-time spectrum for elastic dynamics $\eta =0$. Here, $d=64$,  \smash{$L/L_0=1.1$}, and \smash{$\bar{\sigma} = 0.005$}.}
    \label{fig:spectrum}
\end{figure}

At long times $t/\tau_E \to \infty$, anisotropic diffusion on the constraint sphere leads to an inhomogeneous probability distribution. To characterize the statistical steady state, we consider the power spectrum $\langle \theta_n^2 \rangle$, which reads
\begin{align}
    \langle \theta_n^2 \rangle =  C \frac{1/\gamma_n^{2}}{\sum_k 1/\gamma_k^2} \propto \frac{1}{\left(1 + (n/n_0)^2\right)^2}.
\end{align}  
with a crossover mode number $n_0 = \sqrt{\eta \mu L^2/(\pi^2 B)} = (L/L_0)\sqrt{\mathrm{Pl}}/\pi$: slow-diffusing high-frequency modes are less weighted.
This spectrum tends to become flatter as \smash{$\eta \to \infty$} and the diffusion becoming homogeneous, with a sharper drop-off with $n$ than in the elastic regime  $\mathrm{Pl} \ll 1$: In  the elastic case, we recover the standard result for elastic beams under tension that noise generates a power spectrum peaked at $n=1$, with higher modes decaying as $\langle \theta_n^2 \rangle = D/ (n^2 -1)\sim 1/n^2$ for $n>1$ compared to $n^{-4}$ at intermediate plasticity~\cite{doi_theory_2013}. These results are consistent with the observations of Fig.~\ref{fig:spectrum}. Note that the curves do not perfectly collapse onto each other: this is due to the tension $F$ not vanishing exactly at lower $\mathrm{Pl}$ in the presence of noise.

\section{Derivation of memory criterion for the growing plastica}

We consider a model where growth changes the length of the rod as $L(t) = L_1 e^{gt}$, with $g$ the growth rate, while giving a constant bending rigidity $B$.
In this case, growth generates a baseline tension $T = \mu \dot{C}/(2C) \propto g$, but also leads the elastic coarsening rate $Bq_n^2/\mu \sim 1/L(t)^2$ and spectral noise amplitude $\sqrt{2\sigma/L(t)}$ to decay over the growth period. In what follows, we derive explicit expression for the recall criteria as a function of the plastic rate $\eta$ and geometric parameters.

\subsection{Elastic regime}

\begin{figure}
    \centering
    \includegraphics[scale=1]{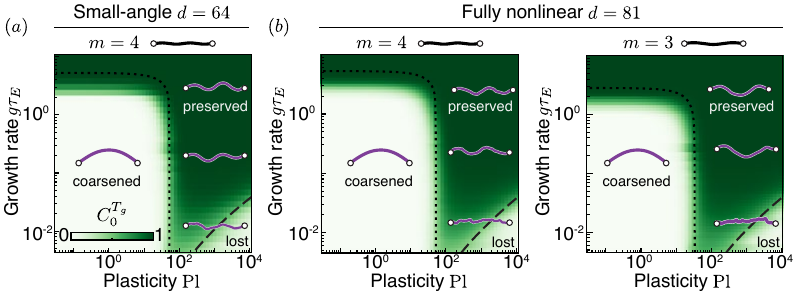}
    \caption{Fully-nonlinear and small-angle simulations both agree with theory in the presence of growth. (a) Small-angle simulations with increased mode number $d=64$ compared to main text Fig. 3. All other parameters identical to Fig. 3a: $L_0 = 1$, $L_1=1.005$, \smash{$L_f/L_1 = 1.1$}, \smash{$\bar{\sigma} = 0.001$} and averaged over $96$ simulations . (b) Fully-nonlinear simulations give similar phenomenology and agreement with theory. $L_0 = 1$, $L_1=1.005$, \smash{$L_f/L_1 = 1.1$}, \smash{$\bar{\sigma} = 0.001$}, results are averaged over $48$ simulations. }
    \label{sfig:vard_growth}
\end{figure}

First, we consider the deterministic criterion for the low plasticity regime. The dynamics for the $n$-th mode $n$ when the $m$-th mode is dominant are now given by
\begin{align}
    \mu\dot{\theta}_n= \left(-Bq_n^2(t) + \frac{\mu \dot{C}}{2C} + B q_m^2(t)  \right)  e^{-\eta t}\theta_n
\end{align}
where now $q_n(t) = \pi n / L(t)$ decreases with time. The mode fractions $r_n = \theta_n^2/C(t)$ then evolve under
\begin{align}
    \dot{r}_n & = 2\frac{\theta_n \dot{\theta}_n}{C} - \frac{\dot{C}}{C} \frac{\theta_n^2}{C} \\
    & = \left(-\frac{2Bq_n^2}{\mu}(t) +  \frac{2B q_m^2(t)}{\mu}  \right) e^{-\eta t} r_n.
\end{align}
The main difference with the no-growth case is the decaying effective bending coefficient as $L(t)$ increases, which adds to the $\eta$-driven vanishing of the growth rate. The total amplification for a finite growth time $T_g = \ln(L_f/L_1)/g$ is then given by
\begin{align}
    \ln \frac{r_n(T_g)}{r_n(0)} \geq \int_0^{T_g} \mathrm{d}t \frac{2B\pi^2}{\mu L_1^2} e^{-(2g+\eta)t} ( m^2 - n^2) & = \frac{2B \pi^2}{\mu L_1^2}\frac{m^2-n^2}{2g + \eta} (1 - e^{-(2g+\eta)T_g})  \\
     & = \frac{2B \pi^2}{\mu L_1^2} \frac{m^2-n^2}{2g + \eta} \left[1 - \left(\frac{L_1}{L_f}\right)^{2+\eta/g}\right] 
\end{align}
To have $r_1(T_g) \leq 1$ for $r_1(0) = \epsilon^2 <1 $ then requires the condition 
\begin{align}
    \frac{2B \pi^2}{\mu L_1^2} \frac{m^2-n^2}{2g + \eta} \left[1 - \left(\frac{L_1}{L_f}\right)^{2+\eta/g}\right] \leq \ln(1/\epsilon^2)
\end{align}
Even when $\eta = 0$, there exists a growth rate $g_0$ such that coarsening is interrupted -- it is  well known that rapid compression can stabilize higher order modes -- with this critical growth rate given by
\begin{align}
    g_0 =  \frac{B\pi^2}{\mu L_1^2 \ln (1/\epsilon^2)}(m^2 - 1) \left(1 -  \frac{L_1^2} {L_f^2}\right).
\end{align}
We find that both the fully nonlinear and small-angle simulations agree with the theory in the presence of growth (Fig.~\ref{sfig:vard_growth}).

\subsection{Large-plasticity regime with growth}

The noise-driven effect persists even with growth tension, after rescaling by $C(t)$. In the presence of noise, in the large plasticity regime we still have 
\begin{align}
    \gamma_n \dot{\theta}_n = F \theta_n + \sqrt{\frac{2\sigma}{L}}\zeta_n(t)
\end{align}
Satisfying $\sum_n \theta_n^2 = C(t)$ now requires
\begin{align}
    \dot{C} = 2 F\sum_n \frac{\theta_n^2}{\gamma_n} = 2 FC \sum_n \frac{r_n}{\gamma_n}
\end{align}
with $r_n = \theta_n^2/C(t)$ the mode fraction, such that $\sum_n r_n =1$ at all times. Solving for the tension gives a result proportional to the mode occupancy averaged mobility
\begin{align}
    F = \frac{\dot{C}}{2C} \frac{1}{\langle \gamma^{-1}\rangle}, \quad \langle \gamma^{-1}\rangle = \sum_n \frac{r_n}{\gamma_n}
\end{align}
We now investigate the deterministic and stochastic dynamics engendered by this effect. 

\subsubsection{Growth in the large-plasticity regime produces limited coarsening}

The growth-induced nonzero tension leads the deterministic dynamics to obey a replicator equation with fitness $1/\gamma_n$,
\begin{align}
    \dot{r}_n & = \frac{2}{C} \theta_n \dot{\theta}_n - \theta_n^2 \frac{\dot{C}}{C^2} \\
    & = \frac{\dot{C}}{C}\left(\frac{1/\gamma_n}{\langle \gamma^{-1}\rangle}-1 \right) r_n
\end{align}
driving the mode fraction to be dominated by slow growing modes. This growth-induced tension thus leads to a similar deterministic coarsening effect as in the elastic limit, but how important is this contribution to coarsening? Near mode-$m$ dominance, $\langle \gamma^{-1}\rangle \approx 1/\gamma_m$, the fundamental mode fraction grows as
\begin{align}
    \frac{\dot{r}_1}{r_1} \approx  \frac{\dot{C}}{C}\left( \frac{\gamma_m}{\gamma_1}-1\right)
\end{align}
The cumulative coarsening due to this deterministic drift is given by 
\begin{align}
    \ln(r_1(T_g)/r_1(0)) &\leq  \int_0^{T_g} \mathrm{d}t \,\frac{\dot{C}}{C} e^{-2g t}\frac{B\pi^2 (m^2-1)}{\mu \eta L_1^2} \nonumber \\ &= \int_0^{T_g} \mathrm{d}t \,\frac{g L_0 e^{-2g t}}{L_1e^{gt}-L_0} \frac{B\pi^2 (m^2-1)}{\mu \eta L_1^2} \approx \frac{B \pi^2(m^2 - 1)}{\mu \eta L_0^2} \left[ -  \ln \left(   1 - \frac{L_0}{L_1}\right) - \left(\frac{L_0}{L_1}\right) - \frac{1}{2}\left(\frac{L_0}{L_1}\right)^2\right]
\end{align}
with the last approximation holding when integrated to infinity. Interestingly, this last result does not depend on the growth rate $g$ nor on the final length. Up to a geometry-dependent factor, this is the same coarsening criterion as in the elastic regime. For $L_1/L_0 \gg 1$, we have to leading cubic order
\begin{align}
    \mathrm{Pl}_c^{\eta} = \frac{\pi^2(m^2-1)}{6\ln(1/\epsilon)} \left(\frac{L_0}{L_1}\right)^3 + O\left(\left[\frac{L_0}{L_1}\right]^4 \right) \approx \frac{1}{6}\mathrm{Pl}^\text{elastic}_c \left(\frac{L_0}{L_1}\right) \to 0
\end{align}
while for $L_1/L_0 = 1.005$ as in our simulations $\mathrm{Pl}_c^\eta \approx 1.9 \mathrm{Pl}_c^\text{elastic}$.
This effect is strongest at small $\eta$ where the validity of the high plasticity approximation is less clear, and it corrects the lower bound on $\eta$ to prevent elastic coarsening. Hence, at high $\eta$ noise-driven mode mixing is still dominant, and this coarsening effect can be safely ignored.

\subsubsection{State diffusion with growth}

The results above indicate that we can still ignore the effect of deterministic coarsening at large plasticity, which is then still dominated by the effect of stochastic mode diffusion. We now compute the timescale of diffusion with $g>0$. In the presence of growth, the diffusion coefficient is now time-dependent as length varies. The total angular drift is now given by
\begin{align}
    \Phi(T_g) = \int_0^{T_g} (d_\text{eff}(t)-1) D_0(t)\mathrm{d}t & \approx \int_0^{T_g}  (d_\text{eff}(0) -1) D_0(t) \mathrm{d}t  \\ & =  \int_0^{T_g} \frac{(d_\text{eff}(0)-1) \sigma}{ \mu^2 \Delta(t)}\mathrm{d}t  = \frac{(d_\text{eff}(0)-1) \sigma}{ \mu^2 g L_0} \ln \left( \frac{(L_f -L_0)L_1}{(L_1 - L_0)L_f}\right)
\end{align}
where $L_1$ is the length at time $t=0$. $\Phi$ has to be $\leq 1$ for preservation of the initial condition. In the large $\eta$ limit, $d_\text{eff} \to d$ from below and there is thus a critical growth rate above which diffusion cannot happen fast enough to lose the initial condition
\begin{align}
    g_\infty = \frac{(d-1) \sigma}{ \mu^2 L_0} \ln \left( \frac{(L_f -L_0)L_1}{(L_1 - L_0)L_f}\right).
\end{align}
Simulations in the small-angle regime quantitatively agree with fully nonlinear simulations and the theory (Fig.~\ref{sfig:vard_growth}).

\section{Alternative model with external viscosity}

Our results above and in the main text assume an internal viscosity of the material. To understand how different damping and dynamics affect the results above, in this section we study the plastica in the case where the dominant cause of dissipation is external viscosity from a surrounding medium. In the small slope limit, where the curve centerline $\mathbf{r}$ can be parametrized as a function $\mathbf{r}(x) = (x, h(x))$ and $\partial_x h = \theta \ll 1$, the overdamped case has linear momentum balance $\mathbf{F}' = ( 0, \mu \dot{h})^T$. We then have $F_x = F$ a constant in $x$ and $\partial_x F_y =  \mu \partial_t h$, which leads to the the dynamical equations
\begin{subequations}
\begin{align}
    \mu \partial_t h & = -B(\partial_x^4 h - \partial_x^2\kappa_0) -F\partial_x^2 h \\
    \partial_t \kappa_0 & = \eta (\partial_x^2h-\kappa_0)
\end{align}
\end{subequations}
subject to the inextensibility constraint 
\begin{align}
    L = \int_0^{L_0} \mathrm{d}x \,\sqrt{1+(\partial_x h)^2} \approx L_0+ \frac{1}{2}\int_0^{L_0} \mathrm{d}x \,(\partial_x h)^2
\end{align}
We again consider hinged (simply supported) boundary conditions such that $h=\partial_{xx}^2h = 0$ at both ends. Expanding in sine modes with $q_n = \pi n/L_0$, we have
\begin{align}
    h(x,t) = \sum_{n=1}^N h_n(t)\sin(q_n x), \quad \kappa_0(x,t) = \sum_{n=1}^N -q_n^2 c_n(t)\sin(q_n x).
\end{align}
In mode space, the inextensibility constraint $(1/2)\int_0^{L_0} \mathrm{d}x\, (\partial_x h)^2 = L - L_0$ becomes
\begin{align}
   \sum_{n=1}^N q_n^2 h_n^2 = 4 \frac{\Delta(t)}{L_0} 
\end{align}
and the mode amplitudes now evolve under
\begin{subequations}
    \begin{align}
        \mu \dot{h}_n & = (-Bq_n^4 +q_n^2F)h_n + B q_n^4 c_n, \\
        \dot{c}_n & = \eta(h_n - c_n).
    \end{align}
\end{subequations}
The tension $F$ can be found by time-differentiating the inextensibility constraint, leading to
\begin{align}
    \frac{1}{\mu}\sum_n q_n^2 \left[ h_n  (-Bq_n^4 +q_n^2 F)h_n + h_n B q_n^4 c_n\right] = \frac{2\dot{\Delta}}{L_0}
\end{align}
which gives 
\begin{align}
    \left( \sum_n q_n^4 h_n^2\right)F  = \mu\frac{2\dot{\Delta}}{L_0} + B \left( \sum_n q_n^6 h_n(h_n-c_n)\right)  
\end{align}
This tension leads to qualitatively similar dynamics, with faster coarsening and an elliptic, rather than spherical, constraint manifold. The criterion for plasticity to prevent elastic coarsening of a pure mode $m$, which has tension $F = Bq_m^2$ can be found by considering the new replicator for the mode $r_n = q_n^2 h_n^2/ C$, now weighted by the wavenumber squared
\begin{align}
    \mu \dot{r}_n = 2Bq_n^2 ( q_m^2 - q_n^2) r_n
\end{align}
which leads to a modified recall criterion
\begin{align}
    \ln(r_1/\epsilon^2) = \int_0^\infty \frac{2B}{\mu}q_1^2(q_m^2 - q_1^2) e^{-\eta t}\mathrm{d}t = \frac{2\pi^4 B}{\mu \eta L_0^4} (m^2-1) \leq \ln(1/\epsilon^2).
\end{align}
This criterion is valid at low noise when the dominant escape channel is towards the fundamental mode. At higher noise, the system can also visit partially coarsened state and stay frozen there, as the transition rates between modes $m\to m-1$ are faster than transitions between $m\to 1$ by a ratio $\rho = (m-1)(2m-1)/(m+1) >1$. Further work could explore the phase diagram of memory retention in the presence of these more complex transients.

\begin{figure}
    \centering
    \includegraphics[scale=1]{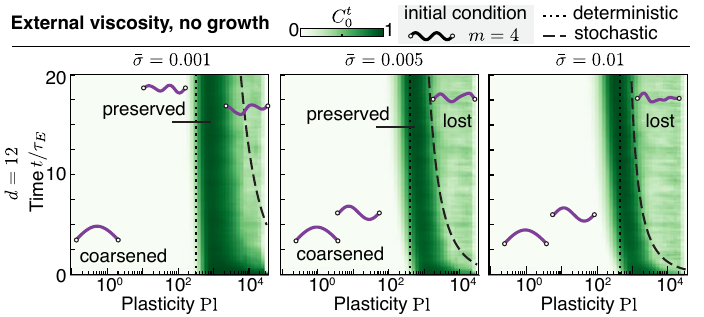}
    \caption{Dynamics of recall for the plastica with external viscosity. We find similar phenomenology as the internal viscosity, with modified thresholds and observe intermediate coarsening for $10\lesssim\mathrm{Pl}\lesssim 10^3$ for the larger noise levels. Here we use the Pearson correlation of $r_n = q_n^2 h_n^2$ to weight spectral components comparably to the internal viscosity case. }
    \label{fig:nogrowthH}
\end{figure}

The large-plasticity stochastic dynamics now read, in the absence of growth, as 
\begin{align}
    (\mu + B q_n^4/\eta) \dot{h}_n = \sqrt{\frac{2\sigma}{L}} \zeta_n
\end{align}
which means that the rescaled mode $u_n = n h_n$ diffuses on a constraint sphere with a modified effective diffusion
\begin{align}
    D_d =(d_\mathrm{eff}-1) D_0, \qquad d_\text{eff} = \sum_n \frac{n^2}{(1+ B q_n^4/(\mu\eta))^2}
\end{align}
with $D_0 = \sigma/(\Delta \mu^2)$. Despite the more complex transients, the resulting criteria yield satisfactory boundaries (Fig.~\ref{fig:nogrowthH}).

\bibliographystyle{unsrt}
\bibliography{biblio}